\documentclass[final,2p,times,twocolumn]{elsarticle}

\usepackage{amssymb}

\usepackage{soul}
\usepackage{graphicx}
\usepackage{dcolumn}
\usepackage{bm}
\usepackage{xcolor}
\usepackage{subfigure}
\usepackage{mathptmx}
\usepackage{amsmath}

\begin{document}

\begin{frontmatter}

\title{Strong Quantization of Current-carrying Electron States in $\delta$-layer Systems}

\author[label1]{Denis Mamaluy}
\author[label1]{Juan P. Mendez}
\affiliation[label1]{organization={Cognitive and Emerging Computing, Sandia National Laboratories},
            addressline={1515 Eubank SE}, 
            city={Albuquerque},
            postcode={87123}, 
            state={NM},
            country={USA}}

\begin{abstract}

We present an open-system quantum-mechanical real-space study of the conductive properties and size quantization in phosphorus $\delta$-layers systems, interesting for their beyond-Moore and quantum computing applications. Recently it has been demonstrated that an open-system quantum mechanical treatment provides a much more accurate match to ARPES measurements in highly-conductive, highly-confined systems than the traditional approaches (i.e. periodic or Dirichlet boundary conditions) and, furthermore, it allows accurate predictions of conductive properties of such systems from the first principles. Here we reveal that quantization effects are strong for device widths $W<10$~nm, and we show, for the first time, that the number of propagating modes determines not only the conductivity, but the distinctive spatial distribution of the current-carrying electron states. For $W>10$~nm, the quantization effects practically vanish and the conductivity tends to the infinitely-wide device's values. 

\end{abstract}


\end{frontmatter}


\section{Introduction}\label{sec:introduction}

Highly conductive $\delta$-layer systems, i.e. thin, high-density layers of dopants in semiconductors are actively used as a platform  for  exploration  of  the  future  quantum  and  classical  computing  when  patterned  in  plane  with  atomic precision. Such structures, with the dopant densities above to the solid solubility limit have been shown to possess very high current densities and thus have a strong potential for beyond-Moore and quantum computing applications. However, at the scale important for these applications, i.e. devices with sub-20 nm physical gate/channel lengths and/or sub-20 nm widths, that could compete with the future CMOS \cite{Spectrum:2021}, the conductive properties of such systems are expected to exhibit a strong influence of size quantization effects.

\begin{figure}[t]
  \centering
  \includegraphics[width=\linewidth]{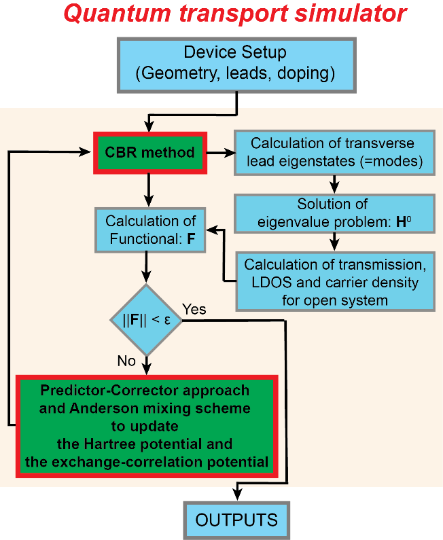} 
  \caption{Flow chart of the charge self-consistent Contact Block Reduction (CBR) method \cite{Mamaluy:2003,Mamaluy:2005,Khan:2007,Mamaluy:2015,Gao:2014}.}
  \label{fig:method}
\end{figure}

Recently it has been demonstrated in \cite{Mamaluy:2021} that to extract accurately the conductive properties of highly-conductive, highly-confined  systems, an open-system quantum-mechanical analysis is necessary, e.g. the Non-Equilibrium Green's Function (NEGF) formalism. In this work we employ an efficient computational implementation of NEGF, which is called Contact Block Reduction (CBR) method \cite{Mamaluy:2003,Mamaluy:2005,Khan:2007,Mamaluy:2015,Gao:2014}, to investigate the conductive properties of Si: P $\delta$-layer systems and their size quantization. A flow-chart of the algorithm implemented in our CBR simulator is shown in Fig.~\ref{fig:method}. For the charge self-consistent solution of the non-linear Poisson equation, we employ a combination of 1) automatic Fermi level determination using Neumann boundary conditions for the non-linear Poisson equation (where the uncertainty by energy is eliminated through the charge self-consistent coupling to the Schrodigner equation with Dirichlet and open-system boundary conditions \cite{Mendez_CS:2022}), 2) the open-system predictor-corrector approach \cite{Khan:2007} and 3) the Anderson mixing scheme \cite{Gao:2014}. 

In this work we demonstrate that quantization effects are strong for device widths $W<10$~nm, whereas, for $W>10$~nm, the quantization effects gradually vanish and the conductivity tends to the infinitely-wide device's values. Similarly, we also show, for the first time, that the number of propagating modes for narrow device widths determines not only the conductivity, but the distinctive spatial distribution of the current-carrying electron states \cite{Mendez_CS:2022}. This existing strong spacial quantization of the current-carrying states could be utilized in novel electronic $\delta$-layer switches.

\section{Results and discussion}\label{sec:results and discussion}

\begin{figure}
  \centering
  \includegraphics[width=\linewidth]{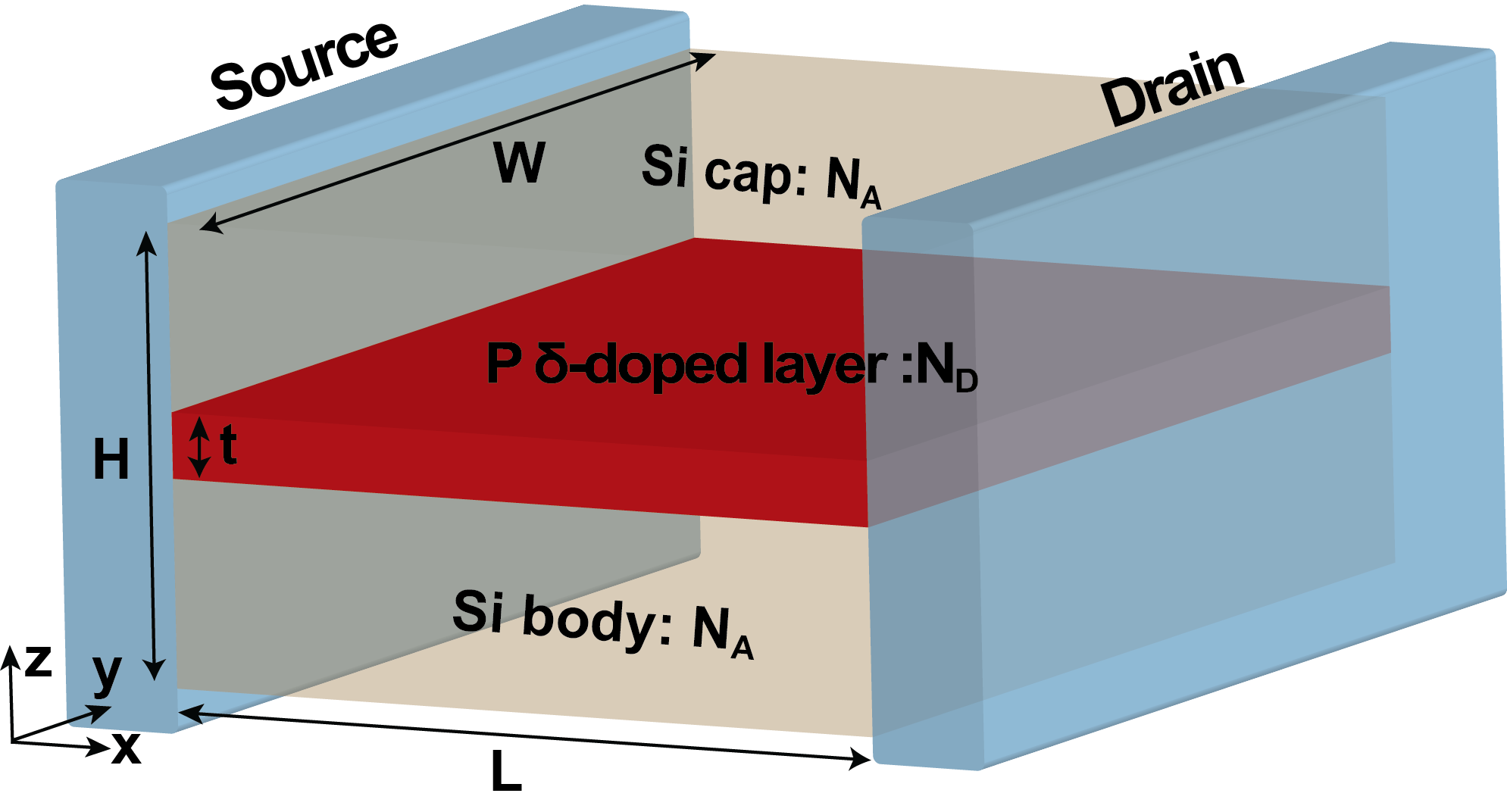} 
  \caption{The Si:P $\delta$-layer device used in our simulations is composed of a Si body, a very high P-doped layer and a Si cap. Open-system quantum-mechanical boundary conditions are applied at the source and drain regions}
  \label{fig:computational_model}
\end{figure}

We have applied our open-system framework to investigate the effects of size quantization in Si: P $\delta$-layer systems on the conductivity at the cryogenic temperature of 4~K. The device is shown in Fig.~\ref{fig:computational_model}, which is composed of a Si body, a very high P-doped layer and a Si cap. As discussed in \cite{Mamaluy:2021}, inelastic scattering is neglected since in Si:P $\delta$-layer systems the phase-relaxation length $l_{\psi}$ is larger than the mean free path $l_m$ at low temperatures \cite{Goh:2006,Mazzola:2014}. However, we also note that all elastic scatterings, including  electron-electron and electron-electric fields interactions, are taken into account in our method through the self-consistent Hartree + LDA exchange-correlation terms. Scattering on discrete charged impurities in $\delta$-layer tunnel junctions was analysed in \cite{Mendez_CP:2021}.

\begin{figure}[t]
  \centering
  \includegraphics[width=\linewidth]{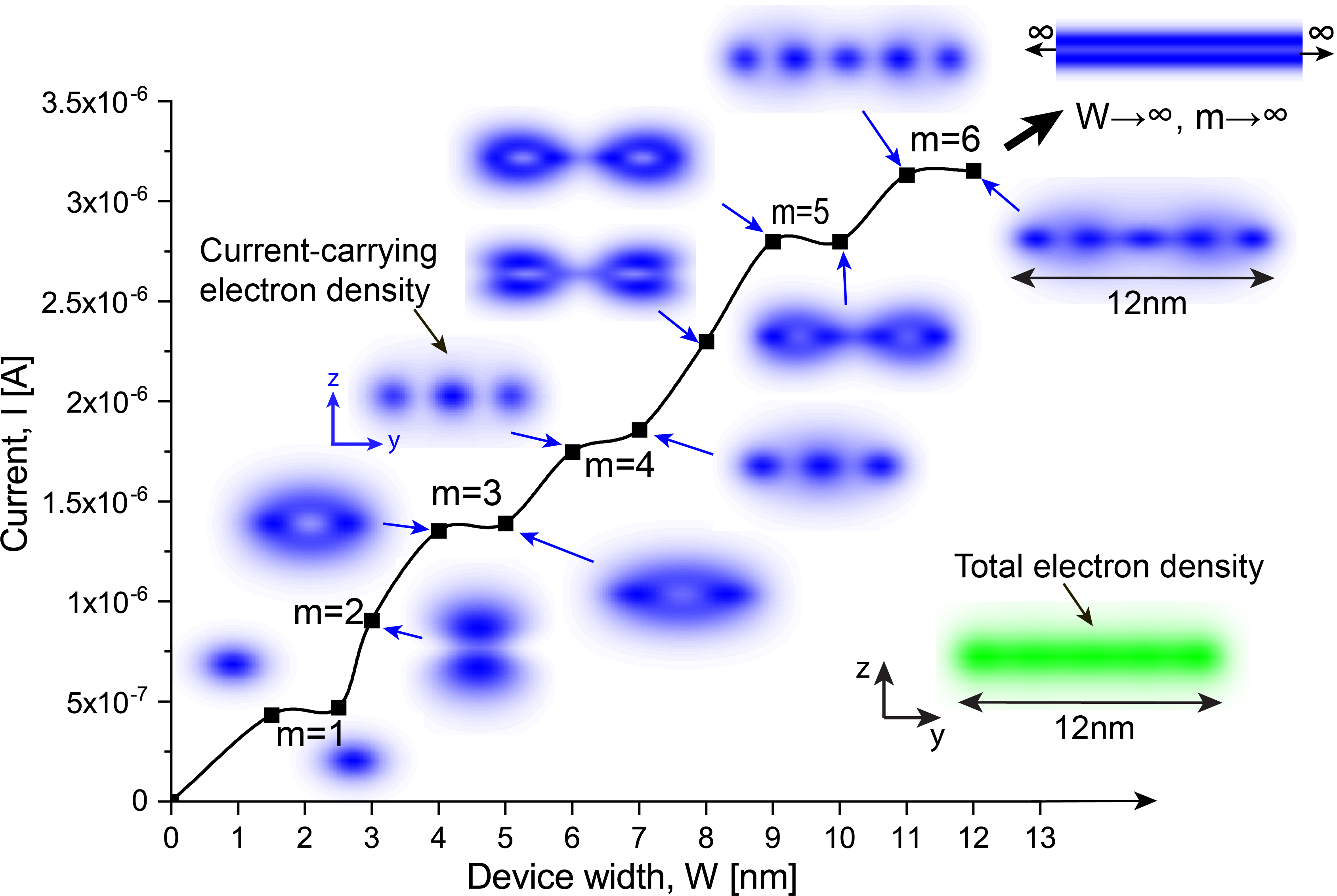} 
  \caption{Propagation modes (m) for Si: P $\delta$-layer systems.  Current $I$ vs device width $W$ for $\delta$-layer systems: the insets in blue color show the spatial distributions of current-carrying modes across a $y-z$ plane, indicating the corresponding number of propagating modes $m$; Inset in green color shows the total electron density that includes all (not just current-carrying) occupied electron states for a device width of $W=12$~nm. For all calculations, $N_D=1.0\times10^{14}$cm$^{-2}$, $N_A=5.0\times10^{17}$~cm$^{-3}$, $t=0.2$~nm and an applied voltage of $1$~mV.}
  \label{fig:propagating modes}
\end{figure}

The corresponding dependence of the current (or, equivalently, the conductance) on the device width $W$ is shown in Fig.~\ref{fig:propagating modes}, taken from our work \cite{Mendez_CS:2022}. We first report that the size quantization effects are strong for device widths $W<10$~nm, whereas, for $W>10$~nm, the quantization effects gradually vanish and the conductivity tends to the infinitely-wide device's values ($W\to\infty$) \cite{Mendez_CS:2022}. The existence of the conduction steps for narrow devices due to each new propagating mode is well known experimentally since 1980’s \cite{Wees:1988}. The total number of propagating modes $m$ depends on the number of peaks in the density of states (DOS) and is mainly determined by three factors: 1) the $\delta$-layer doping level $N_D$, 2) the $\delta$-layer doping thickness $t$, and 3) the device width $W$. Here we report, however, that in highly-confined, highly-conductive $\delta$-layer systems, the quantum number $m$, representing the number of propagating modes, determines not just the total current, but also the spatial distribution of the corresponding current-carrying electrons. 

\begin{figure}[t]
  \centering
  \includegraphics[width=\linewidth]{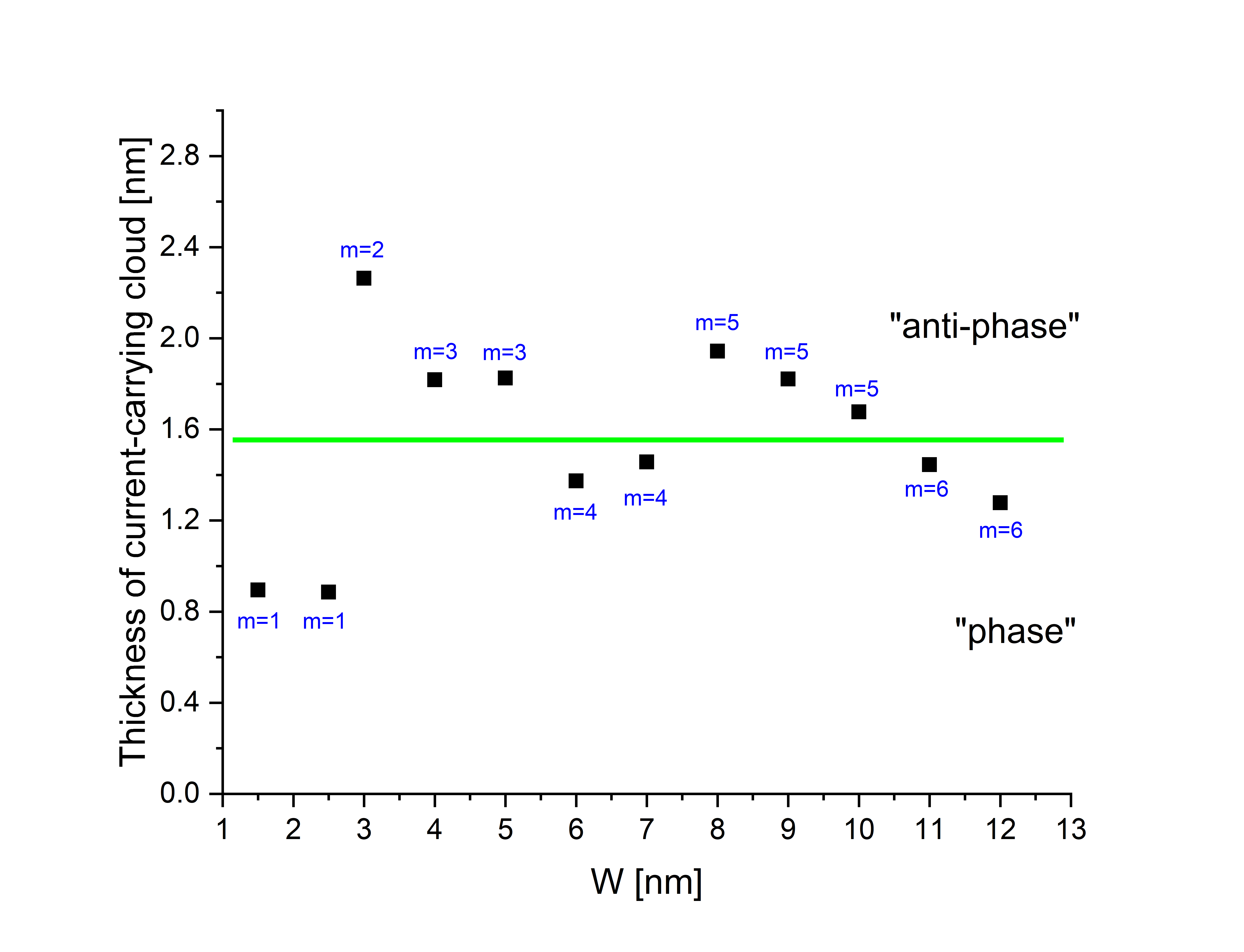} 
  \caption{Average thickness of the \emph{current-carrying electron cloud} in Si: P $\delta$-layer systems vs device width $W$. It demonstrates that "phase" distributions ($m=1,4,6,\dots$) provide more compact placement of current-carrying electrons than "anti-phase" distributions ($m=2,3,5,\dots$). For all calculations, $N_D=1.0\times10^{14}$cm$^{-2}$, $N_A=5.0\times10^{17}$~cm$^{-3}$, $t=0.2$~nm and an applied voltage of $1$~mV.}
  \label{fig:dZ}
\end{figure}

The spatial distribution of the current-carrying electron states, $n_{curr.-carr.}(y,z)$, can be obtained by performing the energy integration of the local density of electron states (LDOS) weighted by the corresponding current spectrum $i(E)$ and normalized with the total current as: $n_{curr.-carr.}(y,z)=\int LDOS(y,z,E)i_e(E)dE/\int i_e(E)dE$. The spatial distribution of current-carrying electrons for different device widths $W$ is shown in Fig.~\ref{fig:propagating modes} as insets in blue color, as well as the corresponding number of propagating mode $m$. Additionally, the total electron density is also included in the figure as an inset in green color, only exhibiting weak spatial quantization along the $y-$direction. We can first discern from these results that not all free electrons contribute to the total current, by comparing the free electron density (inset in green color) and the current-carrying electron density (inset in blue color). Additionally, the specific portion of electrons with energies close to the Fermi level, i.e. the current-carrying states, do exhibit a strong spatial quantization. Indeed, for $m = 1$ the propagating mode reaches the maximum concentration at the center of the structure, the mode that corresponds to $m=2$ is ”excited” into the further penetration along the confinement direction ($z-$axis), leaving the center relatively depopulated (in terms of the current-carrying states), the mode $m=3$ is again "pushed out" of the center along both $z-$ and $y-$ axis, and so on. When $W\to\infty$, the number of propagation modes in $y-$direction becomes infinite $m\to\infty$, as expected. Thus, the modes $m=1$, and $m=4,6$, etc. tend to form a regular "phase" distribution of the current-carrying states (i.e. the states distributed closer to the center of the $\delta$-layer along z-axis), while the modes $m=2$, and $m=3,5$, etc. form ”anti-phase” distributions (i.e. the states distributed further from the center of the $\delta$-layer along z-axis). This is also illustrated in Fig.~\ref{fig:dZ}, showing the average current-carrying electron cloud thickness as a function of device width $W$ and the number of propagating modes $m$. As can be seen, "phase" current-carrying distributions $(m=1,4,6,\dots)$ are located closer to the phosphorus $\delta$-layer, while "anti-phase" current-carrying electrons $(m=3,4,5,\dots)$ are further away from the center of the $\delta$-layer than "phase" distributions. 

Finally, we also note that this strong spacial quantization of the current-carrying states for narrow device widths could be utilized in novel electronic $\delta$-layer switches, where the number of propagating modes and their match/mis-match could be controlled by external electric fields, thus strongly affecting the current. Additionally, in regular $\delta$-layer conductors the particular distribution of current-carrying states directly affects their penetration depth into Si body and cap, which typically has a large concentration of impurities (see e.g. \cite{Ward:2020}). Thus, the control of the number of propagating modes may give an additional degree of control over the rate of impurity scattering.

\section{Conclusions}\label{sec:conclusions}
We employed an efficient computational open-system quantum-mechanical treatment to explore the conductive properties of Si: P $\delta$-layer systems and their size quantization effects for sub-20 nm width $\delta$-layers. We reported a strong spatial quantization of the current-carrying states for device widths $W<10$~nm, which could be utilized in novel electronic $\delta$-layer switches. The number of propagating modes $m$ determines not just the total current, but also the spatial distribution of the corresponding current-carrying electrons. For device widths $W>10$~nm, the quantization effects practically vanish and the conductivity tends to the infinitely-wide device's values.

\small{\emph{This work is funded under Laboratory Directed Research and development (LDRD) program, Project No. 227155, at Sandia National Laboratories. This article has been authored by an employee of National Technology $\&$ Engineering Solutions of Sandia, LLC under Contract No. DE-NA0003525 with the U.S. Department of Energy (DOE). The employee owns all right, title and interest in and to the article and is solely responsible for its contents. The United States Government retains and the publisher, by accepting the article for publication, acknowledges that the United States Government retains a non-exclusive, paid-up, irrevocable, world-wide license to publish or reproduce the published form of this article or allow others to do so, for United States Government purposes. The DOE will provide public access to these results of federally sponsored research in accordance with the DOE Public Access Plan https://www.energy.gov/downloads/doe-public-access-plan. This paper describes objective technical results and analysis. Any subjective views or opinions that might be expressed in the paper do not necessarily represent the views of the U.S. Department of Energy or the United States Government.}}

\bibliographystyle{elsarticle-num} 
\bibliography{references.bib}

\begin{thebibliography}{10}
\expandafter\ifx\csname url\endcsname\relax
  \def\url#1{\texttt{#1}}\fi
\expandafter\ifx\csname urlprefix\endcsname\relax\def\urlprefix{URL }\fi
\expandafter\ifx\csname href\endcsname\relax
  \def\href#1#2{#2} \def\path#1{#1}\fi

\bibitem{Spectrum:2021}
S.~K. Moore, “intel: Back on top by 2025?”, IEEE Spectrum (July 2021).

\bibitem{Mamaluy:2021}
D.~Mamaluy, J.~P. Mendez, X.~Gao, S.~Misra, Revealing quantum effects in highly
  conductive $\delta$-layer systems, Communications Physics 4~(1) (2021) 205.
\newblock \href {https://doi.org/10.1038/s42005-021-00705-1}
  {\path{doi:10.1038/s42005-021-00705-1}}.

\bibitem{Mamaluy:2003}
D.~Mamaluy, M.~Sabathil, P.~Vogl, Efficient method for the calculation of
  ballistic quantum transport, J. Appl. Phys. 93~(8) (2003) 4628--4633.
\newblock \href {https://doi.org/10.1063/1.1560567}
  {\path{doi:10.1063/1.1560567}}.

\bibitem{Mamaluy:2005}
D.~Mamaluy, D.~Vasileska, M.~Sabathil, T.~Zibold, P.~Vogl, Contact block
  reduction method for ballistic transport and carrier densities of open
  nanostructures, Phys. Rev. B 71 (2005) 245321.
\newblock \href {https://doi.org/10.1103/PhysRevB.71.245321}
  {\path{doi:10.1103/PhysRevB.71.245321}}.

\bibitem{Khan:2007}
H.~R. {Khan}, D.~{Mamaluy}, D.~{Vasileska}, Quantum transport simulation of
  experimentally fabricated nano-finfet, IEEE T. Electron Dev. 54~(4) (2007)
  784--796.
\newblock \href {https://doi.org/10.1109/TED.2007.892353}
  {\path{doi:10.1109/TED.2007.892353}}.

\bibitem{Mamaluy:2015}
D.~Mamaluy, X.~Gao, The fundamental downscaling limit of field effect
  transistors, Applied Physics Letters 106~(19) (2015) 193503.
\newblock \href {https://doi.org/10.1063/1.4919871}
  {\path{doi:10.1063/1.4919871}}.

\bibitem{Gao:2014}
X.~Gao, D.~Mamaluy, E.~Nielsen, R.~W. Young, A.~Shirkhorshidian, M.~P. Lilly,
  N.~C. Bishop, M.~S. Carroll, R.~P. Muller, Efficient self-consistent quantum
  transport simulator for quantum devices, J. Appl. Phys. 115~(13) (2014)
  133707.
\newblock \href {https://doi.org/10.1063/1.4870288}
  {\path{doi:10.1063/1.4870288}}.

\bibitem{Mendez_CS:2022}
J.~P. Mendez, D.~Mamaluy, Conductivity and size quantization effects in
  semiconductor $\delta$-layer systems, Scientific Reports 12~(1) (2022) 16397.
\newblock \href {https://doi.org/10.1038/s41598-022-20105-x}
  {\path{doi:10.1038/s41598-022-20105-x}}.

\bibitem{Goh:2006}
K.~E.~J. Goh, L.~Oberbeck, M.~Y. Simmons, A.~R. Hamilton, M.~J. Butcher,
  Influence of doping density on electronic transport in degenerate si:p
  $\ensuremath{\delta}$-doped layers, Phys. Rev. B 73 (2006) 035401.
\newblock \href {https://doi.org/10.1103/PhysRevB.73.035401}
  {\path{doi:10.1103/PhysRevB.73.035401}}.

\bibitem{Mazzola:2014}
F.~Mazzola, C.~M. Polley, J.~A. Miwa, M.~Y. Simmons, J.~W. Wells, Disentangling
  phonon and impurity interactions in $\delta$-doped si(001), Applied Physics
  Letters 104~(17) (2014) 173108.
\newblock \href {https://doi.org/10.1063/1.4874651}
  {\path{doi:10.1063/1.4874651}}.

\bibitem{Mendez_CP:2021}
J.~P. Mendez, S.~Misra, D.~Mamaluy, Influence of imperfections on tunneling
  rate in $\delta$-layer junctions (2022).
\newblock \href {https://doi.org/10.48550/ARXIV.2209.11343}
  {\path{doi:10.48550/ARXIV.2209.11343}}.

\bibitem{Wees:1988}
B.~J. van Wees, H.~van Houten, C.~W.~J. Beenakker, J.~G. Williamson, L.~P.
  Kouwenhoven, D.~van~der Marel, C.~T. Foxon, Quantized conductance of point
  contacts in a two-dimensional electron gas, Phys. Rev. Lett. 60 (1988)
  848--850.
\newblock \href {https://doi.org/10.1103/PhysRevLett.60.848}
  {\path{doi:10.1103/PhysRevLett.60.848}}.

\bibitem{Ward:2020}
D.~Ward, S.~Schmucker, E.~Anderson, E.~Bussmann, L.~Tracy, T.-M. Lu, L.~Maurer,
  A.~Baczewski, D.~Campbell, M.~Marshall, S.~Misra, Atomic precision advanced
  manufacturing for digital electronics, EDFAAO 22~(1) (2020) 4--10.

\end{thebibliography}

\end{document}